\documentclass[sigconf, screen]{acmart}

\usepackage[normalem]{ulem}
\usepackage{makecell}

\AtBeginDocument{%
  }

\setcopyright{none}
\renewcommand\footnotetextcopyrightpermission[1]{}

\acmDOI{}
\acmISBN{}

\acmConference[GI '26]{Graphics Interface 2026}{June 9--12, 2026}{Kitchener-Waterloo, ON, Canada}
\acmBooktitle{Graphics Interface 2026 (GI '26), June 9--12, 2026, Kitchener-Waterloo, ON, Canada}
\acmYear{2026}
\copyrightyear{2026}
\begin{document}

\title{Context-Aware Explanations for Spatialized Document Layouts}



\author{Wei Liu}
\orcid{0009-0009-6340-8912}
\affiliation{%
  \institution{Department of Computer Science, Virginia Tech}
  \city{Blacksburg}
  \state{Virginia}
  \country{USA}
}
\email{wliu3@vt.edu}

\author{John Wenskovitch}
\orcid{0000-0002-0573-6442}
\affiliation{%
  \institution{Department of Computer Science, Virginia Tech}
  \city{Blacksburg}
  \state{Virginia}
  \country{USA}
}
\email{jw87@vt.edu}

\author{Chris North}
\orcid{0000-0002-8786-7103}
\affiliation{%
  \institution{Department of Computer Science, Virginia Tech}
  \city{Blacksburg}
  \state{Virginia}
  \country{USA}
}
\email{north@vt.edu}

\author{Rebecca Faust}
\orcid{0000-0002-7640-1287}
\affiliation{%
  \institution{Department of Computer Science, Tulane University}
  \city{New Orleans}
  \state{Louisiana}
  \country{USA}
}
\email{rfaust1@tulane.edu}


\begin{abstract}
Spatialized document layouts are widely used for exploratory analysis of text corpora, but interpreting the spatial organization of documents and the relationships between regions remains challenging. Existing approaches primarily summarize document content or explain how layouts are generated, providing limited support for understanding spatial relationships within the layout itself. We present CAPE, a context-aware explanation framework that generates natural-language explanations grounded in both document semantics and layout-derived spatial context. CAPE identifies salient spatial patterns (e.g., clusters, subgroups, outliers, and bridging documents) and constructs multi-level contextual representations to guide LLM-based explanation generation. It supports both AI-guided overview and user-driven exploration, with explanations available at multiple levels of detail. We demonstrate CAPE on news and scholarly document layouts and evaluate it in a controlled user study against keyword-based and content-only LLM baselines. Our results suggest that spatially grounded explanations are perceived as more helpful than content-only baselines for interpreting the spatial organization of document layouts.
\end{abstract}

\begin{teaserfigure}
    \centering
    \includegraphics[width=0.90\textwidth]{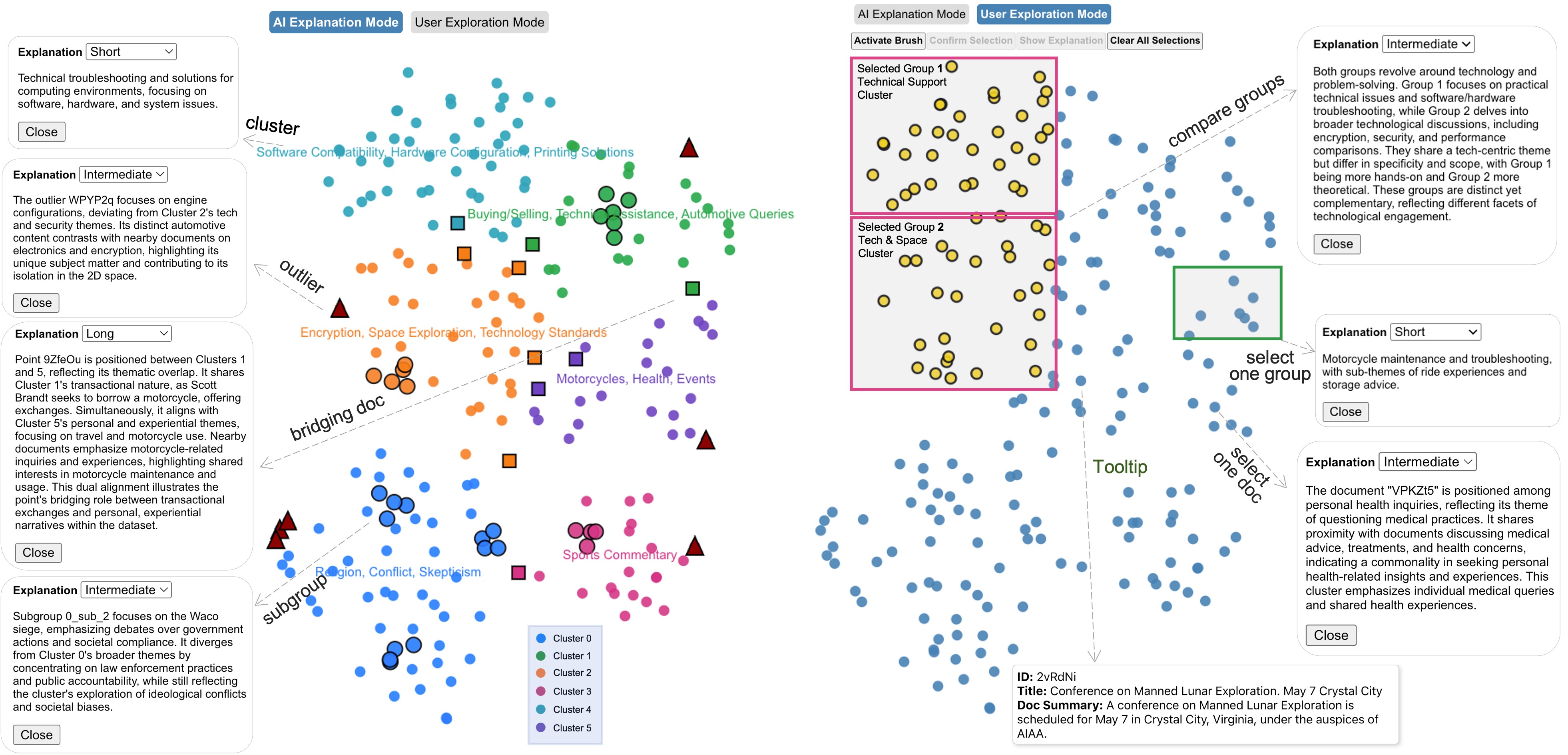}
    \caption{Context-aware explanations for spatialized document layouts, illustrated on a news dataset. In AI Explanation Mode (left), CAPE generates explanations at multiple levels of detail for salient spatial patterns, including clusters, subgroups, outliers, and bridging documents. In User Exploration Mode (right), users can inspect individual documents, request explanations for selected regions, and compare regions. Explanations integrate document semantics with spatial relationships to describe how regions and documents are organized within the layout.}
    \label{fig:cape_ai_mode}
  \end{teaserfigure}




\maketitle

\section{Introduction}
Spatialized layouts of document collections are widely used to support exploratory analysis of text corpora, including workflows such as literature mapping, news exploration, and open-ended corpus sensemaking \cite{liu2024visualizing, endert2013typograph}. By arranging documents in a two-dimensional (2D) space, these layouts allow users to inspect thematic groupings, compare regions, and explore relationships through spatial organization \cite{nguyen2019ten, endertSI, tsai2011dimensionality}. Such representations appear in many forms, including embedding-based projections, manually organized maps, and interactively refined layouts \cite{bian2021deepsi, endertSI, nguyen2019ten, huang2023va+}. Across these settings, users often treat the layout itself as an \textbf{interpretive workspace} for making sense of document collections \cite{bian2021deepsi, german2025narrative, dowling2019interactive}.

Despite their usefulness, spatialized document layouts remain difficult to interpret \cite{patil2023survey, liu2024visualizing}. While the layout makes visible structures such as clusters, outliers, and boundary cases, it does not by itself explain how these structures should be understood in relation to document content. Users may see that documents form groups, that some regions lie close together, or that a document lies between multiple groups, but understanding the meaning of these spatial relationships requires inspecting many documents and mentally integrating local and global context. This process is effortful and time-consuming, especially in large heterogeneous collections.

Existing approaches provide only partial support for this interpretive challenge. Content-oriented techniques such as keyword summaries and topic labels describe what documents or regions are about, but do not explain how they relate within the spatial layout \cite{liu2014topicpanorama, kessler2017scattertext}. Explainable AI methods, in contrast, often focus on how embeddings or projections are generated rather than on how users interpret the resulting spatial representation \cite{bian2021semantic, liu2024visualizing}. More recently, large language models (LLMs) have enabled fluent summaries of document content, but these explanations are typically grounded in text alone and do not account for spatial relationships such as proximity, separation, or intermediate positioning \cite{zhang2024comprehensive}. As a result, current methods offer limited support for \textit{interpreting the spatial organization of document collections}.

In this work, we introduce \textbf{CAPE} (\textbf{Context-Aware Explanations}), a framework for generating natural-language explanations for spatialized document layouts. CAPE treats the current layout as the object of explanation and supports interpretation of its spatial organization, rather than explaining how the layout was produced. To do so, CAPE integrates semantic content with layout-derived spatial context at multiple levels, including global structure, local neighborhoods, and pattern-specific characteristics. This enables CAPE to generate explanations for salient spatial patterns such as clusters, subgroups, outliers, and bridging documents, as well as for user-selected documents and regions. The goal is to help users understand not only what documents or regions are about, but also how they relate to surrounding parts of the layout.

We target analysts and researchers who use spatialized document layouts as exploratory workspaces for understanding document collections, such as news corpora, scholarly literature, and related text collections. This work is guided by two research questions: 

\textbf{(RQ1)} How can natural-language explanations for spatialized document layouts be constructed by integrating semantic information with spatial context?

\textbf{(RQ2)} Does grounding explanations in spatial context help users interpret spatial relationships and organization within a layout, compared with content-only baselines?

To address these questions, we present the CAPE framework, demonstrate it through interactive visualizations, and evaluate it in a controlled user study against keyword-based and content-only LLM baselines. Our results indicate that spatially grounded explanations are perceived as more helpful for interpreting spatial organization and relationships within document layouts than content-only baselines.

This work makes the following contributions:

\begin{itemize} 
    
    \item A context-aware explanation framework for spatialized document layouts that integrates semantic and spatial context.
    
    \item A multi-level explanation strategy supporting both AI-guided overview and user-driven exploration.

    \item An interactive design demonstrating how context-aware explanations can be integrated into exploratory analysis of spatialized document layouts.

    \item Empirical evidence that spatially grounded explanations are perceived as more helpful than content-only baselines for interpreting spatialized document layouts.
\end{itemize}

\section{Related Work}
\label{sec:related_work}
CAPE draws on prior work in spatialized document layouts, explanations of spatial representations, and language-based support for data interpretation.

\textbf{Spatialized Document Layouts}
Spatialized document layouts arrange document collections in a 2D space, where spatial proximity reflects semantic relationships \cite{endertSI, bian2021deepsi}. Prior work has proposed a range of approaches for constructing and interacting with such layouts, including topic model visualizations, embedding-based projections, and interactive document maps \cite{andrewsSpaceToThink, endertSI, nguyen2019ten, dowling2019interactive, bian2021deepsi, inspire1995}. 
These methods allow users to identify clusters, compare regions, and navigate document collections through spatial organization \cite{liu2024visualizing}. To support interpretation, many methods augment layouts with content-oriented summaries such as keyword summaries, topic labels, or representative documents \cite{endert2013typograph, singh2017document, heimerl2014word, nokkaew2018keyphrase, liu2014topicpanorama, ruotsalo2013directing, kessler2017scattertext}. For example, systems such as TopicPanorama \cite{liu2014topicpanorama} and Scattertext \cite{kessler2017scattertext} provide keyword- or phrase-based summaries to help users understand document collections. While these techniques help characterize what documents or regions are about, they primarily focus on semantic content, leaving visible spatial relationships within the layout, such as grouping, separation, and intermediate positioning, largely unexplained.

\textbf{Explaining Spatial Representations}
Explainability has been extensively studied in the context of embeddings, dimensionality reduction, and visual representations \cite{wexler2019if, lundberg2017unified, faust2018dimreader, ribeiro2016should}. Prior work has proposed methods to explain projection behavior, feature contributions, and distortions in low-dimensional embeddings, helping users understand how high-dimensional data are mapped into spatial representations \cite{liu2024visualizing, bian2021semantic}. These approaches are valuable for assessing projection quality and diagnosing model behavior. However, their primary focus is on explaining the computational processes that generate a layout, leaving users to interpret the resulting spatial organization largely on their own. In many exploratory settings, users reason directly about spatial relationships such as proximity, separation, grouping, and overlap, without reference to the underlying model. As a result, methods that explain projection mechanisms do not necessarily support understanding of how documents and regions should be interpreted within the layout. CAPE complements this line of work by focusing on the layout as an interpretive representation and generating explanations grounded in layout-derived spatial context.

\textbf{Language-Based Explanations for Data Interpretation}
Natural language explanations are increasingly used to support data interpretation, including summarizing document collections, labeling clusters, and assisting exploratory analysis workflows \cite{zhang2024comprehensive, sui2024table, choe2024enhancing}. Recent advances in LLMs have enabled fluent and flexible explanations based on textual content, and have been integrated into visualization workflows to support sensemaking \cite{suh2023sensecape, bendeck2024empirical, fu2024scene}. For example, methods such as LangLasso \cite{buchmuller2026langlasso} and TextCluster Explainer \cite{raval2023explain} allow users to obtain textual summaries of selected regions or clusters, improving accessibility and supporting interactive exploration. However, these approaches primarily describe the semantic content of documents or regions and do not explicitly account for their relationships within the spatial organization of the layout. 

\textbf{Summary and Gap} Across these lines of work, existing approaches either focus on summarizing document content or explaining how spatial representations are generated. \emph{Few integrate semantic content with layout-derived spatial context to support understanding of how documents, regions, and spatial patterns relate within the layout}. This gap motivates the design of CAPE.

\section{Design Considerations for Explaining Spatialized Document Layouts}
\label{sec:design_cons}
Spatialized document layouts serve as exploratory workspaces \cite{andrewsSpaceToThink, endert2013typograph} in which users interpret document collections by reasoning about visible structures such as proximity, separation, grouping, and boundary cases \cite{endert2012semantic, andrewsSpaceToThink, smilkov2016embedding}.
Prior work on sensemaking and visual analytics suggests that users construct meaning by iteratively comparing regions, identifying patterns, and relating local observations to global structure \cite{andrews2013impact, endert2012semantics, andrewsSpaceToThink, endert2012semantic}. Interpreting a spatialized layout is therefore a relational and iterative task centered on understanding how documents, regions, and patterns are organized and connected within the layout. 
This perspective motivates four design considerations for explaining spatial organization in document layouts.

\textbf{DC1: Integrating Spatial and Semantic Context.}
Interpretation requires connecting what documents are about with how they are organized in the layout. Spatial proximity is often interpreted as similarity, separation suggests divergence, and intermediate positioning may indicate overlap between themes \cite{andrews2013impact, endert2012semantics, andrewsSpaceToThink, endert2012semantic, smilkov2016embedding}. Explanations that consider only semantic content or only spatial structure provide incomplete support for interpretation.

\textbf{DC2: Explaining Salient Spatial Patterns.}
Users tend to organize their understanding around perceptually salient structures such as clusters, subgroups, outliers, and bridging cases \cite{andrews2013impact, endert2012semantics, andrewsSpaceToThink, endert2012semantic, lisle2021sensemaking, smilkov2016embedding}. These patterns serve as natural units for reasoning about the layout \cite{gorg2012combining, dowling2019interactive, pienta2015scalable, jeon2025stop}. Explanations should therefore target not only individual documents but also higher-level structures and their relationships.

\textbf{DC3: Supporting Multiple Levels of Explanation.}
Exploratory analysis involves moving between high-level overview and detailed inspection \cite{suh2023sensecape, endert2013typograph, heer2012interactive, keim2008visual}. Users often begin with concise summaries to understand global structure, and then require more detailed explanations when examining specific regions or documents. Explanation mechanisms should therefore support multiple levels of detail.

\textbf{DC4: Enabling User-Driven Exploration.}
Sensemaking is iterative and user-driven, with users shifting focus based on emerging questions \cite{pirolli2005sensemaking, keim2008visual, suh2023sensecape}. Users may wish to request explanations for specific documents, compare regions, or examine boundary cases. Explanations should therefore be generated on demand and adapt to user-selected areas of interest.


\section{CAPE Framework}
\label{sec:CAPE_Method}
CAPE is a framework for generating context-aware explanations of spatialized document layouts. Given a 2D layout of documents and their associated textual content, CAPE produces natural-language explanations that integrate semantic meaning with layout-derived spatial relationships.

\begin{figure}[t]
    \centering
    \includegraphics[width=1\linewidth]{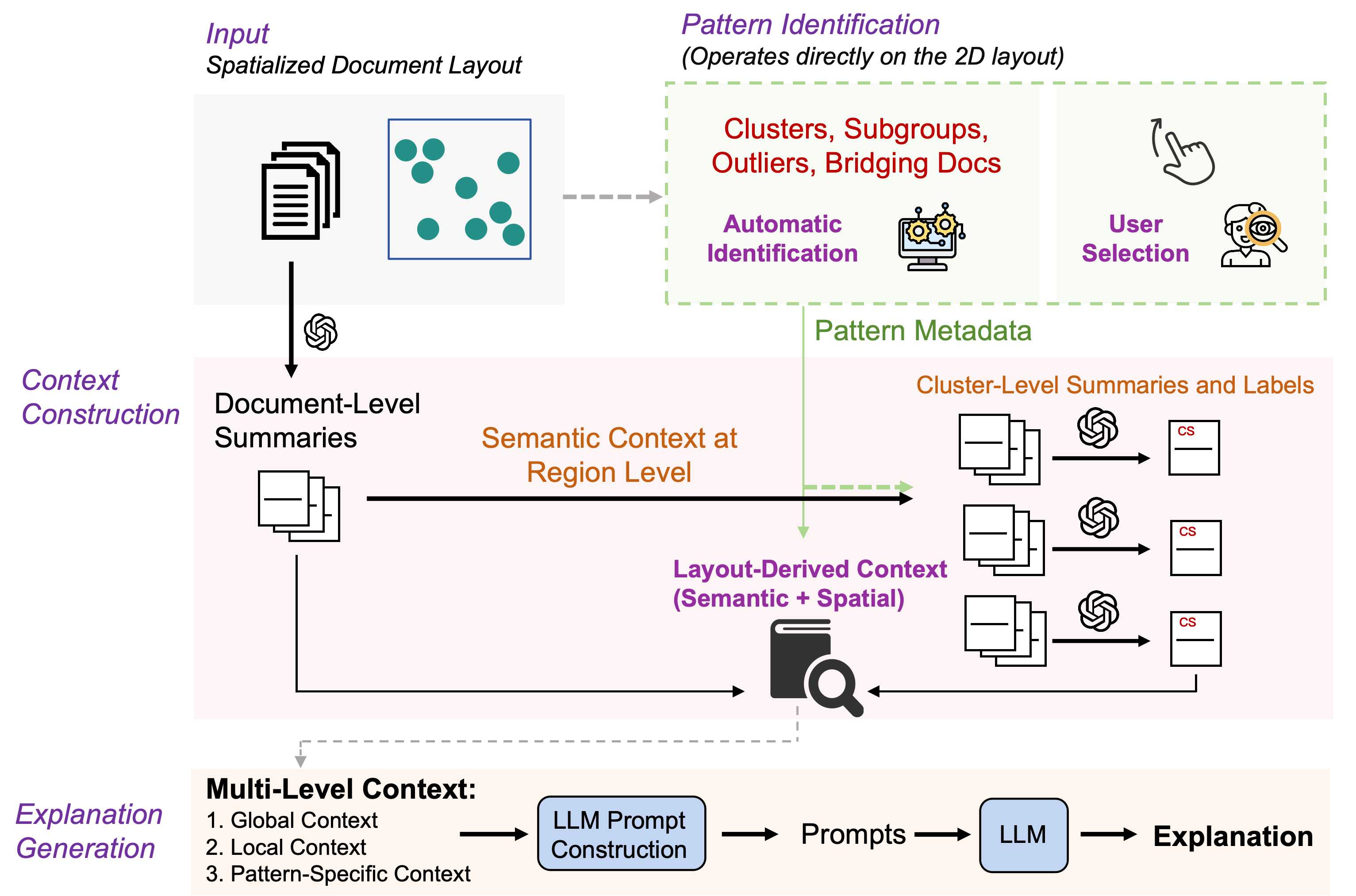}
    \caption{Overview of the CAPE framework. CAPE identifies explanation targets through automatic detection of salient spatial patterns (clusters, subgroups, outliers, and bridging documents) or user selection of documents or regions. It then integrates document semantics with layout-derived spatial context to guide LLM-based explanation generation.}
    \label{fig:cape_pipeline}
\end{figure}

To address \textbf{RQ1} and operationalize the design considerations described in Section~\ref{sec:design_cons}, CAPE follows a structured pipeline consisting of three stages (Figure \ref{fig:cape_pipeline}): (1) \textbf{identifying salient spatial patterns} in the layout (\emph{DC2, DC4}), (2) \textbf{constructing structured contextual representations} that integrate semantic and spatial information (\emph{DC1}), and (3) \textbf{generating explanations} conditioned on the pattern type or user query (\emph{DC2, DC4}), with outputs available at multiple levels of detail (\emph{DC3}).


\subsection{Input: Spatialized Document Layouts}
CAPE operates on a spatialized document layout in which documents are arranged in a 2D space and associated with textual representations. \textbf{The framework is agnostic to how the layout is generated}, and can be applied to embedding-based projections, manually organized maps, or interactively refined layouts. 

CAPE treats the layout as the current representation presented to the user and generates explanations with respect to this representation. CAPE does not assume that spatial proximity strictly reflects high-dimensional similarity. Instead, it focuses on spatial relationships as they appear in the layout, including proximity, separation, and intermediate positioning. For dynamic or interactive layouts, CAPE can be applied to the current state of the layout, with contextual representations recomputed as spatial relationships change.

\subsection{Identifying Salient Spatial Patterns}
\label{sec:pattern_auto_detec}
To support interpretation, CAPE identifies spatial patterns that are likely to draw users’ attention during exploration (\emph{DC2}). These patterns correspond to commonly observed and perceptually salient structures in spatialized document layouts, including:

\begin{itemize}
    \item \textbf{Clusters}: cohesive regions of related documents;
    \item \textbf{Subgroups}: denser local structures within clusters;
    \item \textbf{Outliers}: documents that are spatially separated from surrounding points;
    \item \textbf{Bridging documents}: documents positioned between multiple regions and reflecting potential thematic overlap.
\end{itemize}

This stage is designed to approximate how users visually segment and interpret spatial layouts, not to provide definitive or optimal pattern detection. The pattern identification component is therefore modular and can incorporate different spatial analysis techniques. In our implementation, clusters are approximated using broad
spatial groupings of documents, while subgroups are identified
based on local density within clusters. Outliers are detected based on spatial separation from surrounding regions, and bridging documents are identified through proximity to multiple clusters.

Each identified pattern is associated with basic spatial metadata, including its location in the layout, its relationship to neighboring regions, and the documents it contains and those nearby. This information is not presented directly as algorithmic output, but serves as input for constructing contextual representations used in
explanation generation.

In addition to automatically identified patterns, CAPE supports user-driven selection of documents or regions (\emph{DC4}). Explanations can therefore be generated both for automatically identified structures and for user-defined areas of interest. The framework is selection-agnostic, allowing different interaction techniques such as rectangular or lasso-based selections.

\subsection{Structured Context Construction and Explanation Generation}
\label{sec:context_exp_gen}
To generate explanations that reflect both document semantics
and spatial organization (\emph{DC1}), CAPE constructs a \textbf{structured contextual representation} for each explanation target and uses this representation to guide explanation generation. Details of the contextual representation and the prompt design used for explanation generation are provided in the supplementary materials.

Providing full document collections to a language model is inefficient and can introduce noise. CAPE therefore constructs a compact representation that captures the contextual information most relevant to each spatial pattern or user-defined selection. This enables explanations to remain both informative and scalable, while grounding them in the structure of the layout.

\subsubsection{Context Representation}
\label{sec:context_rep}
CAPE represents each explanation target using a combination of
semantic abstractions and spatial descriptors.

At the \textbf{semantic level}, each document is represented using a concise summary that captures its main idea. These summaries are precomputed and reused across explanation requests. For higher-level structures such as clusters, CAPE derives aggregate representations (e.g., representative summaries) that characterize dominant themes within a region.

At the \textbf{spatial level}, CAPE encodes relationships derived from the layout itself. For each pattern or selected region, this includes:

\begin{itemize}
    \item its position within the layout,
    \item nearby documents in the 2D space, and
    \item relationships to surrounding regions or clusters.
\end{itemize}

Additional spatial descriptors are incorporated depending on the pattern type. For example, outliers are characterized by relative isolation from surrounding regions, while bridging documents are associated with proximity to multiple regions. These semantic and spatial components form a compact contextual representation that captures \textbf{both what documents are about} and \textbf{how they are organized relative to one another in the layout}.

\subsubsection{Multi-Level Context Integration}
To support interpretation at different levels of granularity, CAPE organizes contextual information into three complementary levels:

\begin{itemize} 
    
    \item \textbf{Global context}, which situates the target pattern within the overall layout structure, including region-level summaries and relationships to neighboring regions;
    
    \item \textbf{Local context}, which captures nearby documents and their summaries, reflecting similarity, contrast, and transitions across adjacent areas;

    \item \textbf{Pattern-specific context}, which captures structural characteristics associated with different spatial patterns, such as relative isolation, proximity to multiple regions, or internal relationships within a cluster.

\end{itemize}

This multi-level organization allows CAPE to balance breadth and specificity. Global context provides structural overview, local context captures fine-grained relationships, and pattern-specific context focuses on characteristics that are particularly relevant for interpreting a given pattern or selection.

\subsubsection{Pattern- and Query-Aware Explanation Generation}
Given a structured contextual representation, CAPE generates explanations by conditioning a language model on both the contextual information and the explanatory intent. In our implementation, explanations were generated using the OpenAI API with GPT-4o, with the temperature set to 0.2 \cite{openai_api}. Explanation generation is pattern- and query-aware: CAPE \textbf{tailors prompts to each explanation target}. For example:

\begin{itemize}
    \item For \textbf{clusters}, explanations emphasize shared themes and distinctions from neighboring regions;
    \item For \textbf{subgroups}, explanations describe how local variations relate to the broader cluster;
    \item For \textbf{outliers}, explanations focus on how a document differs from and relates to surrounding regions;
    \item For \textbf{bridging documents}, explanations highlight connections across multiple regions.
\end{itemize}

CAPE also supports user-driven queries. When users select individual documents, explanations relate the document’s content to its surrounding context. When users select groups of documents or regions, CAPE produces summaries of shared properties or comparative explanations that highlight similarities and differences.

This pattern- and query-aware strategy enables explanations to align with users’ analytical intent while remaining grounded in the spatial structure of the layout. 


\subsubsection{Multi-Level Explanation Outputs}
To accommodate different interpretive needs, CAPE produces explanations at three levels of detail (\emph{DC3}): \textbf{short, intermediate, and long}. Short explanations provide concise summaries for rapid orientation. Intermediate explanations include additional contextual detail about nearby regions and spatial relationships. Long explanations further elaborate on semantic variation and broader structural context. All three levels are generated from the same underlying contextual
representation, differing only in the amount of information expressed in the output. This design supports progressive disclosure of information, allowing users to balance readability and informational depth during exploration.

\section{Interaction Design for Context-Aware Explanations}
\label{sec:visual_design}
To operationalize the design considerations described in Section~\ref{sec:design_cons}, we illustrate how CAPE supports context-aware explanations in spatialized document layouts through two complementary interaction modes (Figure \ref{fig:cape_ai_mode}): an AI-guided overview mode (Figure \ref{fig:cape_ai_mode}, left) for rapid orientation and a user-driven exploration mode (Figure \ref{fig:cape_ai_mode}, right) for on-demand analysis. These modes, together with multi-level explanations, enable users to
interpret spatial organization at different stages of exploratory analysis.
Across both modes, users can hover over documents to inspect metadata such as titles and summaries, and click on highlighted patterns or user-defined selections to view their associated explanations.

\subsection{AI-Guided Overview Mode}
The AI-guided overview mode provides an initial interpretive structure by automatically highlighting salient spatial patterns within the layout, including clusters, subgroups, outliers, and bridging documents (\emph{DC2}). This design foregrounds perceptually salient structures during early-stage sensemaking, allowing users to quickly identify meaningful regions and relationships.

Each highlighted pattern is associated with a context-aware explanation that integrates semantic summaries with spatial relationships. These explanations describe both the main themes of the highlighted structure and how it relates to surrounding areas, helping users form an overview of the layout, identify areas of interest, and locate boundary or transitional cases for further analysis.

\subsection{User-Driven Exploration Mode}
The user-driven exploration mode generates explanations in response to user-selected documents or regions, supporting flexible and hypothesis-driven analysis (\emph{DC4}). 

Users can select individual documents to request explanations of how they are positioned within the layout. These explanations relate the document’s content to its surrounding context, including nearby documents and adjacent regions. When users select groups of documents or regions, they can obtain summaries of shared properties or comparative explanations that highlight similarities and differences across selections. This interaction mode supports a range of exploratory tasks, such as interpreting boundary cases, comparing neighboring regions, and examining localized variations. 

\subsection{Multi-Level Explanations Across Modes}
Both interaction modes provide explanation outputs at multiple levels of detail (\emph{DC3}), enabling users to balance rapid orientation with in-depth analysis. 
This progressive disclosure lets users start with concise summaries and expand to richer explanations as needed.

\section{Usage Scenarios}
We present two usage scenarios to illustrate how CAPE supports interpretation of spatialized document layouts in practice.

\subsection{Exploring Themes in News Documents}
\label{sec:news_df}
We first demonstrate CAPE on a spatialized layout of 216 documents sampled from the 20 Newsgroups dataset, covering diverse topics including technology, politics, religion, science, and recreational activities \cite{lang1995newsweeder}. Documents were embedded using OpenAI's text-embedding-3-small model \cite{openai_api} and projected into a 2D space using t-SNE \cite{van2008visualizing}. 
This scenario illustrates how CAPE supports interpretation of (1) global thematic structure, (2) variation within regions, (3) relationships between neighboring regions, and (4) boundary and transitional cases (Figure \ref{fig:cape_ai_mode}).

Using the AI-guided overview mode (Figure \ref{fig:cape_ai_mode} left), CAPE highlights prominent regions and provides cluster-level explanations that summarize dominant themes while situating each region relative to surrounding areas. This supports rapid understanding of the overall structure of the layout and how major thematic areas are organized.

Within a large region centered on religious and ideological discussions (Figure \ref{fig:cape_ai_mode} left-bottom), CAPE reveals finer-grained structure through subgroup-level explanations. These explanations distinguish localized perspectives, such as theological contradictions, societal attitudes, and debates about governmental authority, while relating them to the broader region. This helps users recognize meaningful variation within an otherwise cohesive area.

CAPE enables comparison between nearby regions by generating explanations that highlight differences in thematic focus. For example, two adjacent regions both relate to technology (Figure \ref{fig:cape_ai_mode} right-top), but differ in scope: one emphasizes practical technical support and troubleshooting, while the other focuses on broader topics such as encryption, security, and space exploration. Comparative explanations clarify how these regions are positioned close together while remaining distinct.

CAPE also supports interpretation of boundary cases by relating documents to multiple surrounding regions. For example, one bridging document (Figure \ref{fig:cape_ai_mode} left) discusses borrowing a motorcycle for a trip, and is explained as combining transactional themes (associated with buying and selling activities) and more personal and experiential themes such as travel and motorcycle use, reflecting its intermediate position in the layout.

\subsection{Exploring Trends in Scholarly Collections}
\label{sec:vis_df}
\begin{figure}[t]
    \centering
    \includegraphics[width=\linewidth]{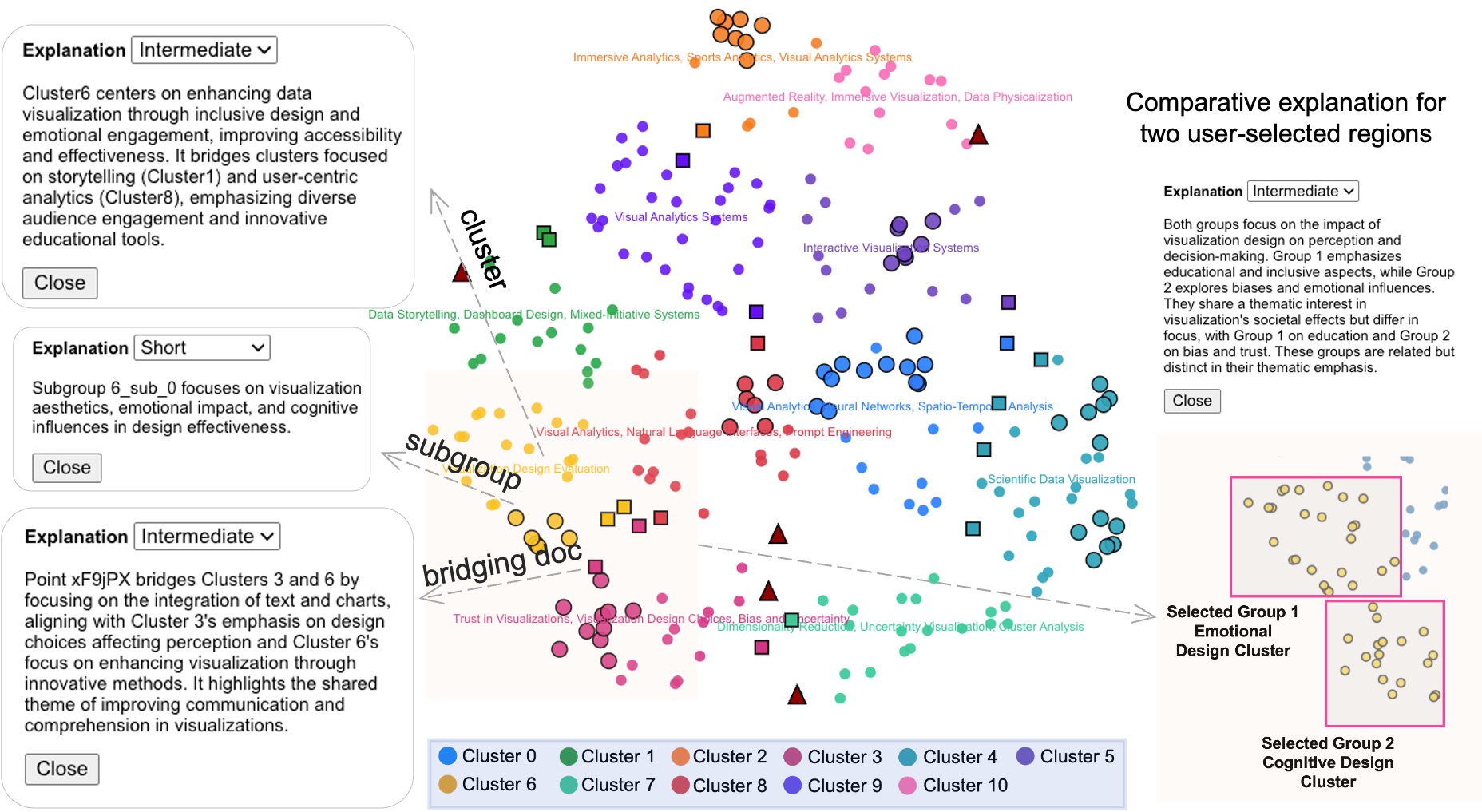}
    \caption{Context-aware explanations on a spatialized layout of IEEE VIS papers. CAPE explains salient spatial patterns identified in the 2D layout, with explanations available at multiple levels of detail (left). Users can also compare selected regions, as illustrated by the contrast between two user-selected design-focused groups (right).}
    \label{fig:user_case_two}
\end{figure}

We next apply CAPE to a spatialized layout of 252 papers from the IEEE VIS 2022 and 2023 proceedings \cite{ieeevis}.
Each paper is represented by its title and abstract, embedded using OpenAI's text-embedding-3-small model \cite{openai_api}, and projected into a 2D space using UMAP \cite{mcinnes2018umap} (Figure \ref{fig:user_case_two}). Compared to the news dataset, this collection reflects a more complex domain in which research themes, methods, and applications are often intertwined.

Using the AI-guided overview mode, CAPE identifies major research areas and provides cluster-level explanations that characterize dominant themes while situating each region within the broader layout. For example, CAPE describes a cluster (Figure \ref{fig:user_case_two} left) centered on inclusive and emotional design in terms of accessibility, engagement, and educational tools, and connects it to neighboring regions focused on storytelling and user-centric analytics.

CAPE further reveals variation within regions through subgroup-level explanations. Within the same cluster, subgroups (Figure \ref{fig:user_case_two} left-middle) highlight more specific thematic emphases, such as visualization aesthetics and cognitive influences on design effectiveness, helping distinguish localized structure within a broader research area.

CAPE also supports comparison between neighboring regions. As shown in Figure \ref{fig:user_case_two} (right), two user-selected groups are both related to visualization design and decision-making, but differ in focus: one centers on educational and inclusive aspects, while the other emphasizes bias, trust, and emotional influences. These explanations clarify how regions can be closely positioned while maintaining distinct thematic orientations.

Finally, CAPE supports interpretation of boundary cases. For example, a bridging document (Figure \ref{fig:user_case_two} left-bottom) is explained as connecting design-oriented concerns about perception and cognition with methodological approaches for improving visualization techniques, highlighting its role in linking adjacent research areas.

\section{User Study: Evaluating Perceived Quality of Explanations}

\begin{table*}[t]
\centering
\small
\caption{Representative example of explanations presented in the study.}
\label{tab:qualitative}
\renewcommand{\arraystretch}{0.9}
\begin{tabular}{p{2cm} p{10.5cm} p{3.6cm}}
\toprule
\textbf{Method} & \textbf{Explanation} & \textbf{Key Characteristics} \\
\midrule

\textbf{CAPE} \newline (Intermediate length)&
``This isolated document focuses on engine configurations, deviating from the broader Technology and Innovation Cluster. Its distinct automotive content contrasts with nearby documents on electronics and encryption, highlighting its unique topic and contributing to its isolation in the 2D space.'' &
Combines semantic and spatial context; explains how the document relates to surrounding regions within the layout. \\
\addlinespace[0.0em]
\cmidrule(lr){1-3}
\addlinespace[0.0em]

\textbf{LLM \newline (Content Only)} \newline (Intermediate length) &
``The document discusses the differences between flat-four (boxer) and inline-four (I4) engine configurations, highlighting the balance and mounting advantages of the flat-four used by Subaru and Porsche. It notes that flat-fours are smoother and more compact, making them suitable for low hoods and rear compartments, while I4s are better for transverse mounting.'' &
Fluent topical summary but does not account for spatial relationships or layout context. \\\addlinespace[0.0em]
\cmidrule(lr){1-3}
\addlinespace[0.0em]

\textbf{Keyword-based} &
``flat, order, crank, second, engines, second order, subaru'' &
Lists frequent terms; no relational context. \\
\bottomrule
\end{tabular}
\end{table*}

To address \textbf{RQ2}, we conducted a controlled user study to evaluate the perceived quality of explanations generated under different conditions.

\subsection{Study Design}
\label{sec:study-design}
We designed a \textbf{within-subjects} study to compare three explanation strategies:

(1) \textbf{CAPE.} Explanations generated using CAPE incorporated spatial context at multiple levels, including global layout structure, local neighborhoods, and pattern-specific information. All CAPE explanations were presented at an intermediate level of detail to ensure comparability.

(2) \textbf{LLM (Content Only).} This baseline used the same language model as CAPE but was provided only with document content summaries, without spatial or neighborhood context. Explanation length was controlled to be comparable to CAPE outputs.
 
(3) \textbf{Keyword-based.} This baseline represents commonly used content-oriented labeling approaches in document visualization. For each spatial pattern, we extracted the top seven TF–IDF \cite{SALTON1988513} keywords from associated documents.

Both CAPE and the LLM (Content Only) baseline used GPT-4o (temperature = 0.2) via the OpenAI API \cite{openai_api}. This design helps isolate the contribution of spatial context from differences in model capability. Table~\ref{tab:qualitative} provides a representative example of the explanations shown to participants under each condition, illustrating how the three strategies differed in the information they conveyed. 

We evaluated explanations across four types of spatial patterns commonly observed in document layouts: clusters, subgroups, outliers, and bridging documents. Two instances of each pattern type were selected from the news layout described in Section~\ref{sec:news_df}, resulting in eight evaluation cases per participant.

\subsection{Participants}
We recruited 10 participants from a university mailing list. The sample included 6 PhD students, 3 master’s students, and 1 undergraduate student. Participants primarily came from computing-related backgrounds. Because background was collected as a multi-select question, categories overlap: 9 participants selected Computer Science, 2 selected Data Science, 2 selected Visualization, and 1 selected Human–Computer Interaction (HCI). Participants reported moderate to high familiarity with data visualization (7-point scale: range = 3–7, median = 5.5). All participants reported at least basic familiarity with interpreting document visualizations. This initial evaluation intentionally recruited participants with sufficient technical background to engage with spatialized document layouts and explanation comparison tasks. The study was approved by the institutional review board.

\subsection{Materials and Procedure}
The study was conducted in person, with each session lasting approximately 25–35 minutes. Participants interacted with a setup that allowed them to view the spatialized document layout and corresponding explanations side by side. When inspecting the layout, participants could view document titles and summaries on demand. Two counterbalanced versions of the study materials were prepared. These versions rotated both (1) the mapping between explanation conditions and labels (A–C) and (2) the order of the eight evaluation patterns. After a brief introduction, participants completed a short tutorial and one practice trial. In each trial, participants were presented with a visualization and three corresponding explanations (one per condition). For each explanation, participants provided ratings on multiple quality dimensions (Section~\ref{sec:us_measures}). After evaluating all three explanations for a given pattern, participants ranked them from most to least helpful. Each participant evaluated eight patterns. After the trials, participants completed a short questionnaire and provided open-ended feedback on helpful and unhelpful aspects of the explanations.

\subsection{Measures}
\label{sec:us_measures}
We collected participants’ subjective ratings using 7-point Likert scales (1 = Not at all, 7 = Very well) across five dimensions.

\textbf{Clarity:} how clear and easy the explanation is to understand.

\textbf{Relevance:} how well the explanation reflects the main themes of the documents or region.

\textbf{Specificity:} how detailed and informative the explanation is.

\textbf{Position-awareness:} how well the explanation helps participants understand how a document or region is positioned
within the layout and how it relates to surrounding areas.

\textbf{Usefulness:} overall perceived helpfulness of the explanation for interpreting the spatial pattern.

We selected clarity, relevance, specificity, and usefulness because they are commonly used criteria in prior work evaluating explanations and visualization support tools, and introduced position-awareness as a study-specific dimension to capture support for understanding spatial relationships within the layout \cite{li2024llms}. In addition to Likert-scale ratings and ranking data, we also collected open-ended responses at the end of the study to capture qualitative feedback about explanation quality.

\subsection{Data Analysis}
All analyses were conducted at the participant level. For each participant, ratings were first averaged across the eight evaluated patterns to obtain a single score per condition and evaluation dimension. This aggregation treated the participant as the unit of analysis in the within-subjects design. Given the ordinal nature of Likert-scale data and the small sample size, we used non-parametric statistical tests. Descriptive statistics are reported as the median and interquartile range (IQR) of participant-level aggregated scores. 
To assess differences between conditions, we conducted Friedman tests for each evaluation dimension. When significant effects were observed, pairwise comparisons were performed using Wilcoxon signed-rank tests with Holm correction for multiple comparisons. Statistical significance was assessed at $\alpha = .05$. 
Effect sizes are reported using Kendall’s $W$ for Friedman tests and rank-biserial correlation ($r_{rb}$) for Wilcoxon tests. For ranking data, participant-level average ranks were computed across the eight evaluated patterns and analyzed using the same procedure (Friedman tests followed by Holm-corrected Wilcoxon tests). Open-ended responses were analyzed using open coding. One researcher first assigned descriptive codes to all responses, after which a second researcher reviewed the coding and emerging themes. Discrepancies were discussed and resolved to consolidate a set of recurring themes.

\subsection{Results}
\label{sec:results}
\subsubsection{Rating Results}
\begin{figure}[t]
  \centering
  \includegraphics[width=\linewidth]{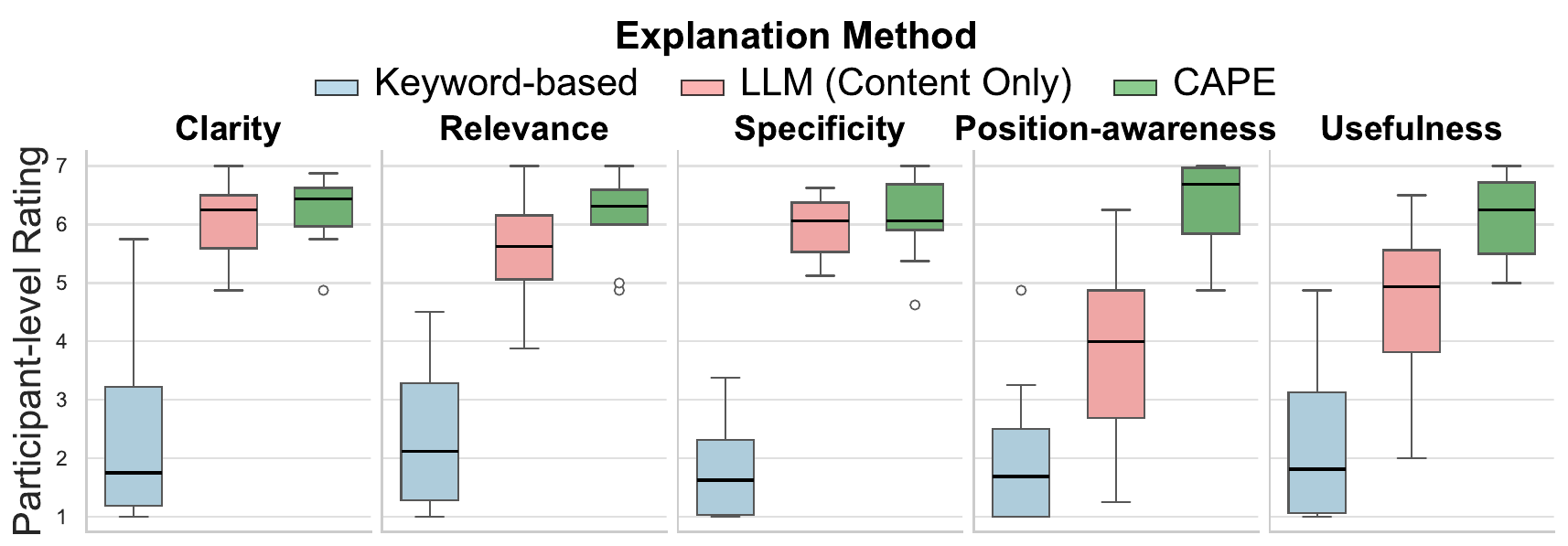}
  \caption{Participant ratings (7-point Likert scale) across explanation methods. CAPE receives higher ratings for position-awareness and usefulness, while ratings for clarity, relevance, and specificity are comparable to the LLM (Content Only) baseline. The keyword-based baseline receives the lowest ratings across all dimensions.}
  \label{fig:mean-ratings}
\end{figure}

Explanation condition had a significant effect on all five evaluation dimensions (Friedman tests, all $p < .01$), with medium to large effect sizes: Clarity ($\chi^2(2)=10.32, p=.0058, W=.52$), Relevance ($\chi^2(2)=15.85, p<.001, W=.79$), Specificity ($\chi^2(2)=15.85, p<.001, W=.79$), Position-awareness ($\chi^2(2)=16.77, p<.001, W=.84$), and Usefulness ($\chi^2(2)=17.59, p<.001, W=.88$). As shown in Figure~\ref{fig:mean-ratings}, CAPE received consistently high ratings across all dimensions, whereas the keyword baseline received low ratings. The LLM (Content Only) condition received ratings similar to those of CAPE for Clarity, Relevance, and Specificity, but lower ratings for Position-awareness and Usefulness.
Holm-corrected Wilcoxon signed-rank tests indicated that both CAPE and LLM (Content Only) were rated significantly higher than the keyword baseline across all dimensions (all $p_{\text{Holm}} < .05$). Comparing the two LLM-based conditions, CAPE received significantly higher ratings for Position-awareness ($W = 1.0, p_{\text{Holm}} = .0078, r_{rb} = .96$) and Usefulness ($W = 3.0, p_{\text{Holm}} = .0152, r_{rb} = .89$), with no significant differences observed for Clarity, Relevance, or Specificity. 
These results suggest that spatial grounding primarily improves perceived support for understanding spatial relationships, while both LLM-based conditions performed similarly on general explanation qualities.

\subsubsection{Ranking Results}

The ranking results were consistent with the rating results. Across all 80 trials (10 participants $\times$ 8 patterns), CAPE was ranked as the most helpful explanation in \textbf{64 out of 80 cases}, followed by LLM (Content Only) in 16 cases, and the Keyword-based baseline in 0 cases. Participant-level analysis showed a significant effect of explanation condition ($\chi^2(2)=19.54, p<.001, W=.98$). Holm-corrected Wilcoxon signed-rank tests indicated that \textbf{CAPE was ranked significantly higher} than both LLM (Content Only) ($W = 0.0, p_{\text{Holm}} = .0071, r_{rb} = 1.00$) and the Keyword-based baseline ($W = 0.0, p_{\text{Holm}} = .0059, r_{rb} = 1.00$). LLM (Content Only) also outperformed the Keyword-based baseline ($W = 0.0, p_{\text{Holm}} = .0059, r_{rb} = 1.00$).

\subsubsection{Qualitative Feedback}
\label{sec:qualitative-feedback}
All participants provided open-ended comments, which offer additional insight into how different explanation strategies supported interpretation. Open coding produced four themes that organize the discussion below.

\begin{itemize}
    \item \textit{Combining semantic summaries with spatial reasoning}: connecting what a document or region is about with its spatial context in the layout.
    
    \item \textit{Understanding relationships to nearby documents and regions}: describing how a document or pattern relates to neighbors.
    
    \item \textit{Clarity and level of detail}: expressing the main idea clearly and at an appropriate level of detail.
    
    \item \textit{Limitations of keyword-based explanations}: keyword-only 
    explanations perceived as too vague to convey meaning or to clarify relationships within the layout.
\end{itemize}

\textbf{Combining Semantic Summaries with Spatial Reasoning.}
Many participants described helpful explanations as integrating both topical summaries and spatial context (7/10). For example, a participant commented that helpful explanations ``talk about the document and also why a document is placed where it's placed'' (P01), while another emphasized explanations that ``summarized both the main topic and why it was positioned in a specific location'' (P02).

\textbf{Understanding Relationships to Nearby Documents and Regions.}
Several participants highlighted the importance of describing relationships to nearby documents or clusters (5/10). For example, a participant noted that helpful explanations should ``[show] their relation to nearby documents'' (P04). Another stated that ``the document-cluster relationship[s] are what is difficult to find by hand'' (P06). These responses suggest that spatially grounded explanations may make such relationships easier to interpret.

\textbf{Clarity and Level of Detail.} Participants also valued explanations that were clear and concise. Some participants described helpful explanations as easy to read and well-structured (4/10), such as ``short but capturing the main theme with easy words'' (P07) and ``simple and only used jargon if necessary'' (P10).  One participant also noted that overly long explanations could be difficult to process, requiring them to ``go back and forth and consume all information'' (P09). These responses suggest a need to balance informativeness with readability.

\textbf{Limitations of Keyword-Based Explanations.} Participants frequently described keyword-based explanations as vague and difficult to interpret (7/10). For example, one participant noted that they ``only gave several words... I have to guess the topic'' (P02), while another described them as ``completely unhelpful'' because ``the missing sentences made it hard to interpret the words properly'' (P05). In addition, several participants mentioned that unhelpful explanations often failed to clarify relationships, making them less useful for understanding spatial organization (7/10).

\subsubsection{Summary of Findings}

Across ratings, rankings, and qualitative feedback, CAPE was consistently preferred over the keyword baseline and was most frequently ranked as the most helpful explanation. Compared to the content-only LLM baseline, CAPE was associated with higher perceived position-awareness and usefulness, while maintaining comparable clarity, relevance, and specificity. These findings suggest that incorporating spatial context supports users in interpreting how documents and regions relate within the layout.

\section{Discussion}
\hspace{1em} \textbf{Explaining Spatial Layouts as Interpretive Representations.} 
Spatialized document layouts are often treated as outputs of embedding and projection pipelines, with interpretability work focusing on how such layouts are generated \cite{huang2023va+, liu2024visualizing, jeon2025stop}. In contrast, CAPE treats the layout itself as an \emph{interpretive representation that users reason about during exploratory analysis}. In this setting, users often attend to visible spatial cues such as grouping, separation, and intermediate positioning, regardless of how the layout was produced. CAPE therefore complements prior work on explainable embeddings and dimensionality reduction by focusing on user-facing interpretation of visible spatial organization.

\textbf{What Spatial Context Adds Beyond Content-Only Explanation.}
Our results suggest that the main contribution of spatial context lies not in improving linguistic quality, but in improving support for reasoning about relationships within the layout. CAPE and the content-only LLM baseline received comparable ratings for clarity, relevance, and specificity, but CAPE was rated higher for position-awareness and overall usefulness. 
This pattern suggests that content-only explanations are effective for summarizing topical content, while spatially grounded explanations provide \emph{additional support for understanding documents and regions in relation to surrounding spatial context}. This is consistent with prior work describing sensemaking in spatial layouts as a relational process centered on comparison, proximity, separation, and transition \cite{endert2012semantic, andrewsSpaceToThink, smilkov2016embedding}.

\textbf{Implications for Explanation Design.}
These findings suggest three implications for explanation design in spatial analytic workflows. First, explanations should explicitly address relationships between documents and regions, not just summarize topical content. Second, explanations may benefit from adapting to different spatial pattern types, since clusters, subgroups, outliers, and bridging cases involve different interpretive questions. Third, explanation mechanisms should support both rapid overview and on-demand exploration, ideally with multiple levels of detail, so that users can move between broad orientation and focused inspection. 

\textbf{Scope and Future Directions.}
CAPE is designed to explain layouts as presented to users, not to validate projection faithfulness or correct distortions introduced during layout generation. This makes the framework applicable across embedding-based, manually organized, and interactively refined layouts, but also means that explanations inherit biases present in the current representation \cite{nguyen2019ten, endertSI, tsai2011dimensionality, liu2026llm}. Future work could explore adaptive explanation mechanisms, integration of additional metadata such as temporal or categorical information, and extension to other forms of spatialized analytic representations.

\textbf{Limitations.}
This work has several limitations. First, the user study involved a relatively small sample (N = 10) drawn primarily from computing-related backgrounds, which limits generalizability beyond technically experienced users. We also did not collect detailed demographic information such as gender; prior work suggests that individual differences can influence visualization interpretation \cite{liu2020survey}. Larger and more diverse samples would help examine whether explanation preferences vary across user groups. Second, because the evaluation focused on subjective ratings and rankings, the findings speak to perceived explanation quality rather than analytical accuracy or efficiency. Third, the study examined two document collections, and additional evaluations would be needed to assess broader domain generality. Finally, CAPE explains layouts as presented and relies on heuristic pattern identification, so explanations may inherit biases or distortions from the underlying layout and may not capture all meaningful structures in more complex settings.

\section{Conclusion}
We presented CAPE, a context-aware explanation framework for spatialized document layouts that supports interpretation of spatial organization rather than explanation of how layouts are generated. By integrating semantic information with layout-derived spatial context, CAPE generates natural-language explanations that help users understand how documents and regions relate within a layout. A controlled user study suggests that spatially grounded explanations are perceived as more helpful for understanding the layout compared with content-only baselines. These findings highlight the importance of incorporating spatial context into explanation mechanisms for spatial representations and suggest opportunities for supporting interpretation in broader exploratory analysis workflows.

\section*{Use of Generative AI}
During the preparation of this work, the authors used ChatGPT to improve writing. After using this tool, the authors reviewed and edited the content as needed and take full responsibility for the content of the publication.



\bibliographystyle{ACM-Reference-Format}
\bibliography{references}


\end{document}